\documentclass[twocolumn]{aastex63}

\graphicspath{figures/}

\usepackage{amsmath}

\newcommand{\result}[1]{{\color{black}#1}}

\newcommand{\maya}[1]{{\color{black}{#1}}}

\usepackage{xspace} 
\newcommand{\Msun}{\ensuremath{\,\mathrm{M_\odot}}\xspace} 
\newcommand{\Gyr}{\ensuremath{\,\mathrm{Gyr}}\xspace} 

\begin{document}

\title{LIGO-Virgo-KAGRA's Oldest Black Holes: \\
Probing star formation at cosmic noon with GWTC-3
}
\author{Maya Fishbach}
\email{fishbach@cita.utoronto.ca}
\affiliation{Canadian Institute for Theoretical Astrophysics, David A. Dunlap Department of
Astronomy and Astrophysics, and Department of Physics, 60 St George St, University of Toronto, Toronto, ON M5S 3H8, Canada}
\author{Lieke van Son}
\affiliation{Center for Astrophysics | Harvard \& Smithsonian, 60 Garden Street, Cambridge, MA 02138, USA}
\affiliation{Anton Pannekoek Institute for Astronomy and GRAPPA, University of Amsterdam, NL-1090 GE Amsterdam, The Netherlands} 
\affiliation{Max-Planck-Institut f\"ur Astrophysik, Karl-Schwarzschild-Straße 1, 85741 Garching, Germany}

\begin{abstract}
In their third observing run, the LIGO-Virgo-KAGRA gravitational-wave (GW) observatory was sensitive to binary black hole (BBH) mergers out to redshifts $z_\mathrm{merge}\approx1$.
Because GWs are inefficient at shrinking the binary orbit, some of these BBH systems likely experienced long delay times $\tau$ between the formation of their progenitor stars at $z_\mathrm{form}$ and their GW merger at $z_\mathrm{merge}$. 
In fact, the distribution of delay times predicted by isolated binary evolution resembles a power law $p(\tau)\propto\tau^{\alpha_\tau}$ with slope $-1\lesssim\alpha_\tau\lesssim-0.35$ and a minimum delay time of $\tau_\mathrm{min}=10$ Myr. 
We use these predicted delay time distributions to infer the formation redshifts of the $\sim70$ BBH events reported in the third GW transient catalog GWTC-3 and the formation rate of BBH progenitors.
For our default $\alpha_\tau=-1$ delay time distribution, we find that GWTC-3 contains at least one system (with 90\% credibility) that formed earlier than \result{$z_\mathrm{form}>4.4$}.
Comparing our inferred BBH progenitor formation rate to the star formation rate (SFR), we find that at $z_\mathrm{form}=4$, the number of BBH progenitor systems formed per stellar mass was \result{$6.4^{+9.4}_{-5.5}\times10^{-6}\Msun^{-1}$} and this yield dropped to \result{$0.134^{+1.6}_{-0.127}\times10^{-6}\Msun^{-1}$} by $z_\mathrm{form}=0$.
We discuss implications of this measurement for the cosmic metallicity evolution, finding that for typical assumptions about the metallicity-dependence of the BBH yield, the average metallicity at $z_\mathrm{form}=4$ was \result{$\langle\log_{10}(Z/Z_\odot)\rangle=-0.3^{+0.3}_{-0.4}$}, although the inferred metallicity can vary by a factor of $\approx3$ for different assumptions about the BBH yield. 
Our results highlight the promise of current GW observatories to probe high-redshift star formation.
\end{abstract}

\section{Introduction}
\label{sec:introduction}
The gravitational-wave (GW) detector network consisting of Advanced LIGO~\citep{2015CQGra..32g4001L}, Advanced Virgo~\citep{2015CQGra..32b4001A} and KAGRA~\citep{2021PTEP.2021eA101A} has observed GW radiation from binary black holes (BBHs) that merge at redshifts $z_\mathrm{merge} \lesssim 1$~\citep[e.g.][]{2019PhRvX...9c1040A,2021PhRvX..11b1053A,2021arXiv211103606T,2021ApJ...922...76N,2022PhRvD.106d3009O}, with planned detector upgrades expanding the network sensitivity to $z_\mathrm{merge} \lesssim 2$~\citep{2015PhRvD..91f2005M,2018LRR....21....3A,2023arXiv230609234W}. Observing stellar-mass binaries that merge beyond $z_\mathrm{merge}\gtrsim2$, an era known as ``cosmic noon" when the Universe formed most of its stars, requires a next generation of GW detectors proposed for the 2030s, such as Einstein Telescope~\citep{2020JCAP...03..050M} and Cosmic Explorer~\citep{2021arXiv210909882E}. These next-generation observatories would be sensitive to compact binary mergers past $z_\mathrm{merge} > 50$~\citep{2019CQGra..36v5002H,2021arXiv211106990K,2022arXiv220211048B}, likely observing every BBH merger in the Universe and opening up enormous discovery potential for learning about high-redshift star formation, the first generation (Pop III) stars, the assembly and growth of the first galaxies, cosmic metallicity evolution, and the formation histories of BBH systems~\citep{2019ApJ...886L...1V,2021ApJ...913L...5N,2022ApJ...933L..41N,2022arXiv220610622C}.

Although the current LIGO-Virgo-KAGRA (LVK) network cannot observe mergers that happened earlier than $z_\mathrm{merge}\approx1$, many of the observed low-redshift mergers may have formed at significantly higher redshifts. Merging binaries experience a delay time between the formation of the progenitor stars and the binary compact object merger. Because massive stars are short-lived, the bulk of this delay time typically consists of the GW-driven binary inspiral. The GW inspiral time $\tau_\mathrm{inspiral}$ is very sensitive to the orbital separation $a$, scaling as $\tau_\mathrm{inspiral} \propto a^4$ \citep{1964PhRv..136.1224P}. In fact, a binary needs to reach extremely short separations $\mathcal{O}(10R_\odot)$ just to merge within the age of the Universe~\citep{1964PhRv..136.1224P}.
Small increases in the orbital separation drastically increase the GW inspiral time, implying that even a narrow distribution of initial orbital separations will cause the distribution of delay times to extend out to very long delays.

\begin{figure}
    \centering
    \includegraphics[width= 0.49\textwidth]{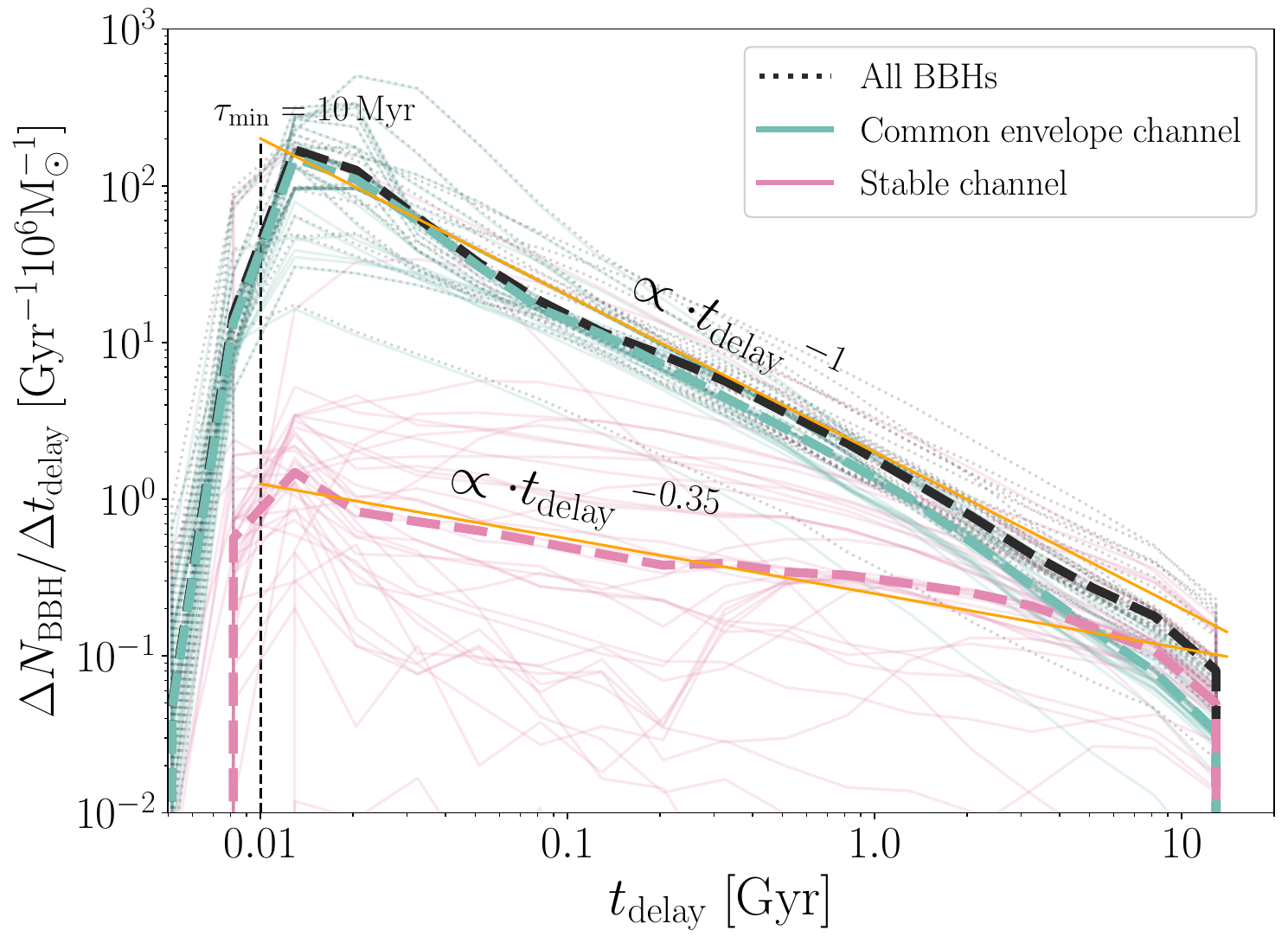} 
    \caption{\label{fig:delayTimes} Delay time distributions derived from all  physics variations explored in \protect{\cite{2022ApJ...940..184V}}. We show histograms of the delay times per $10^{6}\Msun$ of stellar mass formed, assuming a flat-in-log distribution of metallicities, and bin sizes $\Delta \log t_{\rm{delay} } = 0.2$. Thick dashed lines show the median for each formation channel, orange thin lines show the power laws used in this work.}
\end{figure}

{Indeed, different BBH formation channels~\citep[see e.g.,][for a review]{2022PhR...955....1M, 2020FrASS...7...38M} typically predict delay time distributions with a long tail that extends beyond a Hubble time. This applies to dynamically-assembled BBH systems or stellar triples in various environments~\citep[e.g.][]{2016PhRvD..93h4029R,2017ApJ...841...77A,2020MNRAS.498..495D,2020ApJ...896..138Y,2022ApJ...936..184M} as well as those resulting from isolated binary evolution~\citep[e.g.][]{2013ApJ...779...72D,2016MNRAS.463L..31L,2017MNRAS.472.2422M}. Here we focus on predictions from isolated binary evolution.}

{Different isolated binary formation channels generally predict distinct distributions for the separations at BBH formation, and thus distinct delay time distributions. For example, the common envelope channel leads to shorter final separations than the stable mass transfer channel, because common envelopes shrink the binary orbit more efficiently than stable mass transfer \citep[see Fig.~\ref{fig:delayTimes}, and e.g.,][]{2021A&A...647A.153B,2021ApJ...922..110G,2022ApJ...931...17V}.\footnote{This difference is almost `by design' since the energy formalism typically used in population synthesis codes to approximate common envelopes was first formulated to explain very short period double white dwarfs as the progenitors of SN Ia \citep{1984ApJ...277..355W,1984ApJS...54..335I}.}
Similarly, chemically homogeneously evolving stars are expected to stay compact and tend towards shorter delay time distributions {at sufficiently low metallicities} \citep[e.g.,][]{2016A&A...588A..50M,2020MNRAS.499.5941D,2021MNRAS.505..663R}.
}


Because of these distinct predictions, directly measuring the delay time distribution probes the formation channels of GW sources. For mergers involving neutron stars, the delay time distribution can be inferred from a population of their host galaxies~\citep{2019ApJ...878L..13S,2020ApJ...905...21A}, the host galaxy properties of short gamma ray bursts~\citep{2022ApJ...940L..18Z}, or the r-process enrichment history~\citep[e.g.,][]{2022ApJ...926L..36N}. 
For BBH sources without uniquely identified host galaxies, measuring the redshift evolution of the merger rate~\citep{2018ApJ...863L..41F,2020ApJ...896L..32C,2019ApJ...886L...1V,2019ApJ...878L..13S,2023arXiv230207289C,2023ApJ...946...16E} can inform the delay time distribution~\citep{2019ApJ...886L...1V,2021ApJ...914L..30F,2023MNRAS.523.4539K}. \citet{2021ApJ...914L..30F} found that the relatively steep redshift evolution of the BBH merger rate between $z_\mathrm{merge} = 0$ and $z_\mathrm{merge} = 1$, compared to the low-metallicity SFR model from \citet{2017ApJ...840...39M}, favors short delay times, ruling out delay time distributions with typical delays $\gtrsim3$ Gyr. In the absence of a direct model for the SFR,~\citet{2022ApJ...937L..27M} showed that the distribution of delay times between star formation and BBH mergers can be inferred by cross-correlating the redshift distributions of BBH mergers with electromagnetic tracers of star formation (e.g., line intensity mapping). 

Conversely, if the delay time distribution is known, the redshift evolution of the BBH merger population directly constrains the BBH progenitor formation rate. The progenitor formation rate depends on the SFR, stellar initial mass function (IMF), and the cosmic metallicity as a function of redshift. These are highly important yet uncertain processes, particularly at high redshifts~\citep{2014ARA&A..52..415M,2020ARA&A..58..577S,2019A&ARv..27....3M}. Therefore, GW mergers present an exciting opportunity to probe star-forming conditions in the high-redshift Universe~\citep[see, e.g.][for a review]{2022arXiv220610622C}.

Previous studies have studied the BBH progenitor formation rate within a population synthesis framework, simulating BBH merger events for a range of parameters that govern the metallicity-specific SFR. By comparing to LVK observations, one can place constraints on the input parameters~\citep[e.g.][]{2023arXiv230300508R}. Alternatively, one can simultaneously fit the delay time distribution and the progenitor formation rate to the GW observations in a data-driven approach, but the two will be highly degenerate with each other unless one restricts the model flexibility by placing some physical priors~\citep[e.g.][]{2019ApJ...886L...1V}. 

In this work, we assume a simplified form of the delay time distribution motivated by theoretical predictions, and use it to propagate the observed merger redshift of each BBH event in GWTC-3 backward to a probability distribution of its progenitor's formation redshift $z_\mathrm{form}$, {while simultaneously inferring the progenitor formation rate}. 
Our approach sits between a population synthesis forward model and a data-driven inference, and can be thought of as a highly simplified version of ``backward population synthesis"~\citep{2018ApJS..237....1A,2021ApJ...914L..32A,2022arXiv220604062W}. Unlike full backward population synthesis, 
we apply only a couple of predictions from population synthesis and combine them with GW observations, starting with a delay time distribution to infer BBH progenitor formation redshifts. Indeed, one application of the backward population synthesis approach is providing a straightforward consistency check that a given population synthesis simulation predicts delay times and metallicity-specific BBH rates that match both GW and SFR observations. 

{The remainder of this paper is structured as follows. In \S\ref{sec:intuition}, we introduce the theoretically-motivated delay time distributions and derive the relationship between delay times, merger redshifts and formation redshifts. In \S\ref{sec:Rf-inference}, we model the BBH merger rate in terms of a fixed delay time distribution and an unknown progenitor formation rate, and infer the progenitor formation rate from the GWTC-3 data. We then adopt a metallicity-dependent yield $dN_\mathrm{BBH} / dM_\mathrm{SF}(Z)$ motivated by population synthesis and a SFR model from galaxy observations to turn our inference of the BBH progenitor rate into a constraint on the cosmic metallicity evolution (\S\ref{sec:metallicity}).
In \S\ref{sec:zform-pe}, we discuss what our population fit implies for the merger and formation redshifts of individual GWTC-3 BBH events, showing that GWTC-3 likely contains several systems that formed before cosmic noon. 
We discuss limitations of our method and future directions in \S\ref{sec:discussion} before concluding in \S\ref{sec:conclusion}.}

\section{Delay time distributions}
\label{sec:intuition}
{Before inferring the BBH formation rate in the following sections, we first review the relationship between the delay time distribution $p(\tau)$, the merger rate $R_m$ and the formation rate $R_f$, and explain our choice of delay time distribution models.}
Assuming a delay time $\tau$, a binary that forms at lookback time $t_\mathrm{form}$ merges at the lookback time:
\begin{equation}
\label{eq:tform}
t_\mathrm{merge} = t_\mathrm{form} - \tau. 
\end{equation}
{Assuming a distribution of delay times $p_d(\tau)$, a binary that formed at $t_\mathrm{form}$ will merge at a time $t_\mathrm{merge}$ drawn from the distribution:
\begin{equation}
\label{eq:ptm}
p(t_\mathrm{merge} | t_\mathrm{form}) = p_d(\tau = t_\mathrm{form} - t_\mathrm{merge}).
\end{equation}
The rate of mergers at a given time $t_\mathrm{merge}$ depends on the formation rate in addition to the delay time distribution~\citep[e.g.][]{2007PhR...442..166N,2016PhRvL.116m1102A}:
\begin{align}
    \label{eq:Rm-time}
    R_m(t_\mathrm{merge}) &= \int p(t_\mathrm{merge} \mid t_\mathrm{form}) R_f(t_\mathrm{form}) d t_\mathrm{form} \nonumber \\
    &= \int p_d(\tau) R_f(t_\mathrm{form} = t_\mathrm{merge} + \tau) d \tau,
\end{align}
where in the second line we changed variables from $t_\mathrm{form}$ to $\tau$ using Eq.~\ref{eq:ptm}.
Meanwhile, by Bayes theorem, given a binary that merged at time $t_\mathrm{merge}$, the probability it formed at time $t_\mathrm{form}$ is given by:
\begin{equation}
\label{eq:ptf}
p(t_\mathrm{form} | t_\mathrm{merge}) = \frac{p(t_\mathrm{merge}| t_\mathrm{form}) R_f(t_\mathrm{form})}{R_m(t_\mathrm{merge})}.
\end{equation}
Therefore, the BBH merger rate, which we can measure with GW observations, informs a combination of the progenitor formation rate and the delay time distribution (Eq.~\ref{eq:Rm-time}) and allows us to infer the formation times of individual BBH events from their merger times (Eq.~\ref{eq:ptf}).}

Although the above equations are expressed in terms of lookback time, we assume a cosmological model (in this case, flat $\Lambda$CDM with parameters from~\citealt{2016A&A...594A..13P}) to convert between lookback time $t$ and redshift $z$~\citep{2018AJ....156..123A}, so that $z_\mathrm{merge}$ ($z_\mathrm{form}$) is the cosmological redshift at the lookback time $t_\mathrm{merge}$ ($t_\mathrm{form}$).

We assume a power-law parameterization for the delay time distribution with slope $\alpha_\tau$, minimum delay time $\tau_\mathrm{min}$, and maximum delay time 
$\tau_\mathrm{max}$:
\begin{equation}
\label{eq:delay-pl}
p_d(\tau) \propto \tau^{\alpha_\tau}\Theta( \tau_\mathrm{min} < \tau < \tau_\mathrm{max}),
\end{equation}
where $\Theta$ denotes the indicator function.
In reality, the delay time distribution varies depending on the BBH mass and formation metallicity. For simplicity, we neglect this dependence and assign the delay time distribution of Eq.~\ref{eq:delay-pl} to every BBH system regardless of its mass and any assumptions about its formation metallicity, {but see \S\ref{sec:discussion} for a discussion on this assumption.}
We bound our uncertainty in the delay time distribution by varying the power-law slope $\alpha_\tau$.

In Fig.~\ref{fig:delayTimes} we show variations of the delay time distribution inferred from the population synthesis predictions of \cite{2022ApJ...940..184V}. {These results assume a universal \cite{2001MNRAS.322..231K} initial mass function (IMF) and average over a flat-in-log distribution of metallicities.} {Motivated by these models, we explore three slopes: $\alpha_\tau = -1$ (our default model; similar to the prediction for ``All BBHs”, which is typically dominated by the common envelope channel), $\alpha_\tau = -0.35$ (similar to the prediction from the stable mass transfer channel), and an intermediate slope of $\alpha_\tau = -0.7$.}
We further fix $\tau_\mathrm{min} = 10$ Myr {to represent the typical lifetime of a massive star}, and fix $\tau_\mathrm{max}$ to the maximum lookback time of progenitor formation, which we take to be $t_0 = 13.62$ Gyr (corresponding to a maximum progenitor formation redshift $z_0 = 20$). Choices of $\tau_\mathrm{max} > t_0$ are equivalent to changing the normalization of the progenitor formation rate. As we will see in the following, we are not yet sensitive to the choice of $z_0$, as GWTC-3 is unlikely to contain any events that formed earlier than $z_\mathrm{form} \gtrsim 10$ for these choices of delay time distributions.

\section{progenitor formation histories}
\label{sec:Rf-inference}

\begin{figure}
    \centering
    \includegraphics{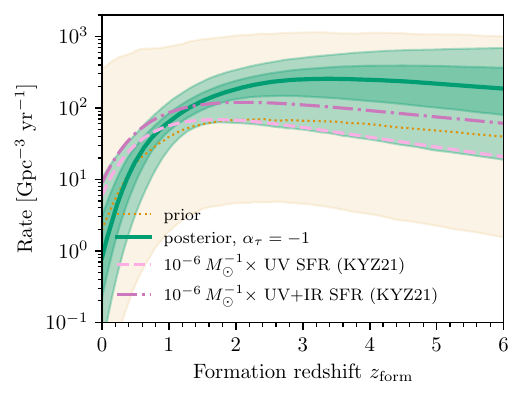}
    \caption{Inferred formation rate assuming the default $\alpha_\tau = -1$ delay time distribution (green line shows median a posteriori formation rate as a function of redshift, while shaded bands enclose 50\% and 90\% posterior probability). We adopt the formation rate parametrization from~\citet{2021ApJ...919...88K} with two parameters controlling the shape and one parameter controlling the amplitude (see Eq.~\ref{eq:Rf-KYZ}); the prior is shown in yellow (dotted line shows median a priori formation rate, while shaded band encloses 90\% prior probability). For comparison, we show the UV SFR (dashed light pink line) and UV+IR SFR (dot-dashed dark pink line) from~\citet{2021ApJ...919...88K} normalized by $10^{-6}\Msun$.}
    \label{fig:rate-form-v-z}
\end{figure}

We now use the GWTC-3 observations to fit the BBH merger rate, which we model as a convolution of an unknown progenitor formation rate and a theoretically-motivated delay time distribution (Eq.~\ref{eq:Rm-time}). The result is a measurement of the progenitor formation rate as a function of redshift (i.e., the formation rate of systems that will merge as BBHs within $\tau_\mathrm{max} = 13.62\Gyr$). 

Our approach is similar to the progenitor formation rate calculation by~\citet{2021ApJ...914L..30F} with GWTC-2 (see their Fig. 7). However, while~\citet{2021ApJ...914L..30F} assumed that the progenitor formation rate follows the low-metallicity SFR given by~\citet{2017ApJ...840...39M} and fit for the metallicity threshold and scatter in the mean metallicity-redshift relation, here we adopt a more agnostic model, simply assuming that the progenitor formation rate can be described by the functional form from~\citet{2021ApJ...919...88K} with free parameters $\mathcal{N}$, $a$, $b$:
\begin{equation}
\label{eq:Rf-KYZ}
R_f(z_\mathrm{form}) = \mathcal{N} T(z_\mathrm{form})^{a} \exp(-b T(z_\mathrm{form})),
\end{equation}
where $\mathcal{N}$ is a normalization in units of Gpc$^{-3}$ yr$^{-1}$, $T(z)$ is the age (in Gyr) of the Universe at redshift $z$, $b$ has units of inverse Gyr, and $a$ is unitless. Although this function naturally falls to zero as $z_\mathrm{form}$ approaches infinity (when $T = 0$), we set $R_f$ to zero for $z_\mathrm{form} > z_0$ with $z_0 = 20$. 
For each of the three delay time distributions we consider, we fit $\mathcal{N}$, $a$ and $b$, along with the BBH mass and spin distribution in a hierarchical Bayesian framework~\citep{2004AIPC..735..195L,2019MNRAS.486.1086M}. Additional analysis details can be found in Appendix~\ref{sec:appendix}.

Fig.~\ref{fig:rate-form-v-z} shows the fit to the progenitor formation history (green) assuming the default $\alpha_\tau = -1$ delay time model. Although we fit the formation rate out to $z_0 = 20$, we only plot the formation rate for $z_\mathrm{form} \leq 6$, because we expect the GWTC-3 events to have formed within this redshift range (see \S\ref{sec:zform-pe} and Fig.~\ref{fig:EDF_zmerge_zform_GWTC3}). For reference, we show the prior in yellow, the~\citet{2021ApJ...919...88K} UV fit to the SFR as the dashed light pink line, and the~\citet{2021ApJ...919...88K} UV+IR fit to the SFR as the dot-dashed dark pink line. Both SFRs are normalized by $10^{-6}\Msun$. 

In the following, we use the UV SFR as the reference SFR because~\citet{2021ApJ...919...88K} find that it better matches the observed stellar mass density evolution, whereas they note that the UV+IR SFR overestimates the observed stellar mass density by up to $\sim0.5$ dex. However, we caution that observational uncertainties affect both the SFR and the stellar mass density~\citep[e.g.][]{2023arXiv230610118N}. These uncertainties are currently not significant compared to uncertainties associated with the GW inference and the choice of delay time model, so we adopt a fixed SFR model for the following proof-of-principle calculations. 

\begin{figure}
    \centering
\includegraphics{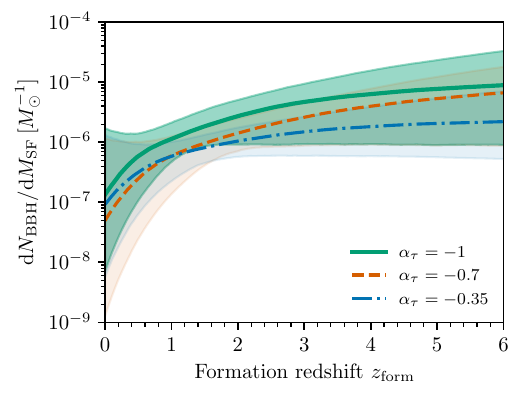}
    \caption{Number of merging BBH systems formed per star-forming mass as a function of redshift $z_\mathrm{form}$, as inferred under three different delay time distributions with power-law slopes $\alpha_\tau = -1$ (default, in green/ solid), $\alpha_\tau = -0.7$ (orange/ dashed) and $\alpha_\tau = -0.35$ (blue/ dot-dashed) and {the~\citet{2021ApJ...919...88K} UV SFR}. Lines denote the median BBH yield at each formation redshift while shaded bands enclose 90\% posterior probability.}
    \label{fig:yield-v-redshift}
\end{figure}

{We find that the inferred BBH progenitor formation rate rises more steeply with increasing redshift than the SFR, in particular over the range $z_\mathrm{form} < 4$ where we expect to get the most meaningful constraints with GWTC-3.}
This is illustrated in Fig.~\ref{fig:yield-v-redshift}, where we show the yield $dN_\mathrm{BBH}/ dM_\mathrm{SF}$ inferred under the three different delay time models. The yield is defined as the number of BBH progenitors that will merge within $\tau_\mathrm{max} = 13.62$ Gyr formed per stellar mass. For power-law delay time distributions with slopes $\alpha_\tau = -1$ or shallower, we infer that the BBH yield decreases with decreasing $z_\mathrm{form}$, from \result{$6.4^{+9.4}_{-5.5}\times10^{-6}\Msun^{-1}$} at $z_\mathrm{form} = 4$ to \result{$0.134^{+1.6}_{-0.127}\times10^{-6}\Msun^{-1}$} at $z_\mathrm{form} = 0$ under the default $\alpha_\tau = -1$ delay time model. 

We find that compared to the default $\alpha_\tau = -1$ delay time distribution, the shallower $\alpha_\tau = -0.7$ and $\alpha_\tau = -0.35$ models provide a worse fit to the GW data in combination with our prior on the progenitor formation history. Under these shallow delay time models, our assumed formation rate model \emph{a priori} limits the merger rate to evolve only by a factor of $\approx 2$ between $z_\mathrm{merge} = 0$ and $z_\mathrm{merge} = 1$, whereas the data prefer a factor of $\approx7$~\citep{2023PhRvX..13a1048A}. This worse fit is further illustrated by the maximum likelihood values of the fits under the different delay time assumptions, which differ by a factor of \result{26.5 (3.3)} in favor of the $\alpha_\tau = -1$ assumption relative to the $\alpha_\tau = -0.35$ ($\alpha_\tau = -0.7)$ assumption. Either our prior for the progenitor formation history is not sufficiently flexible (as we discuss in \S\ref{sec:discussion}), or, if our prior captures the range of reasonable BBH progenitor formation rates, the $\alpha_\tau = -1$ model is a better description of the true BBH delay time distribution.

\section{metallicity evolution}
\label{sec:metallicity}

We can use our measurement of the BBH yield $dN_\mathrm{BBH}/dM_\mathrm{SF}(z_\mathrm{form})$ presented in Fig.~\ref{fig:yield-v-redshift} to infer the cosmic metallicity evolution with redshift. This is because we expect the BBH yield $dN_\mathrm{BBH}/dM_\mathrm{SF}$ to depend strongly on the formation metallicity $Z$~\citep{2010ApJ...715L.138B}. Given a theoretically-motivated assumption for $dN_\mathrm{BBH}/dM_\mathrm{SF}(Z)$, we can use our measurement of $dN_\mathrm{BBH}/dM_\mathrm{SF}(z_\mathrm{form})$ to infer the metallicity distribution as a function of redshift $p(Z \mid z_\mathrm{form})$. These quantities are related as:  
\begin{align}
\label{eq:eta-of-redshift}
    dN_\mathrm{BBH}/&dM_\mathrm{SF}(z_\mathrm{form}) = \nonumber \\ &\int dN_\mathrm{BBH}/dM_\mathrm{SF}(Z)p(Z \mid z_\mathrm{form}) dZ.
\end{align}

\begin{figure}
    \centering
    \includegraphics[width= 0.49\textwidth]{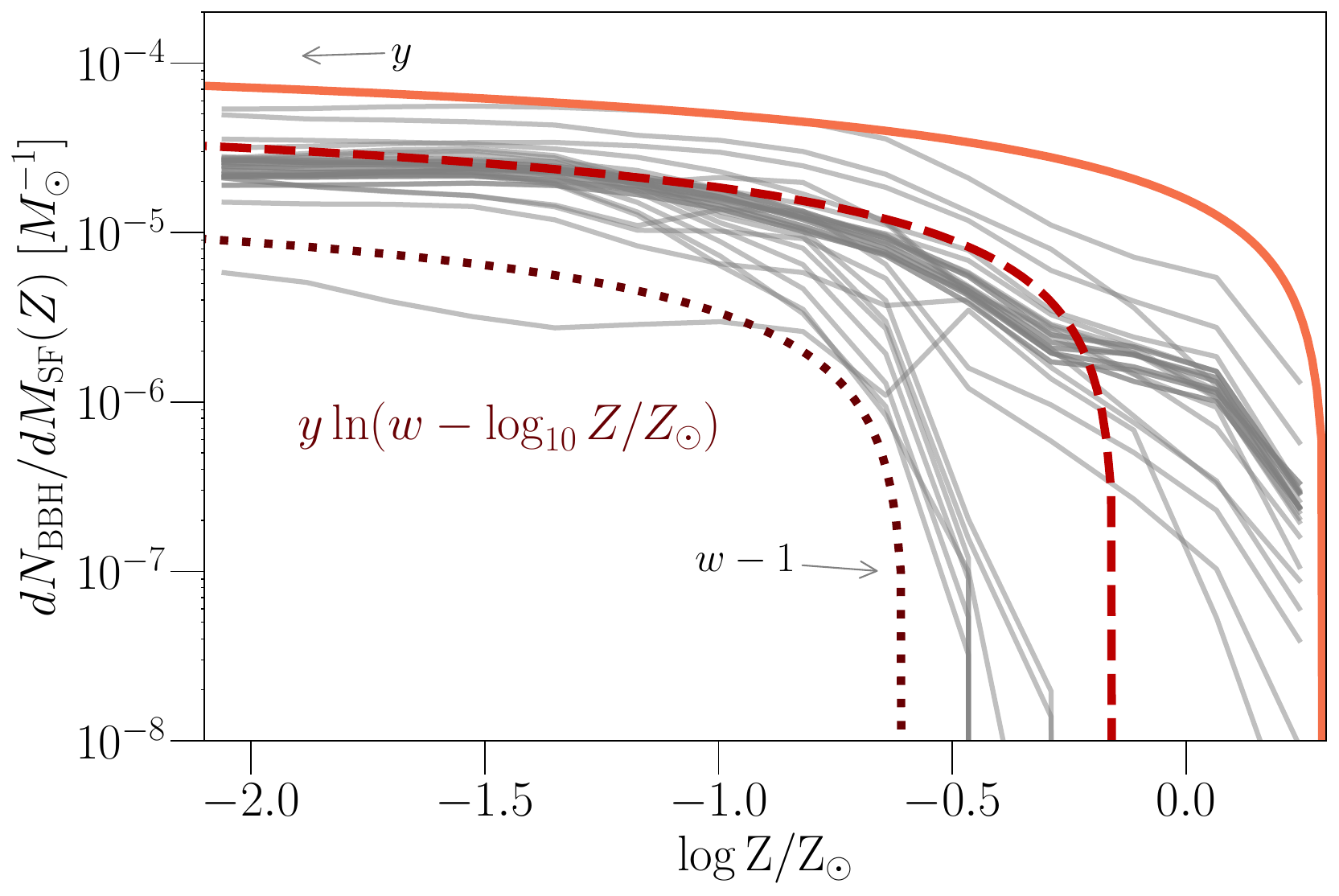}
    \caption{Number of merging BBH systems formed per
star-forming mass as a function of metallicity (i.e., the BBH yield) for the physics variations explored in \citet{2022ApJ...940..184V} -- gray lines. We show analytical approximations of low (dotted) medium (dashed) and high (solid line) BBH yields following Eq. \ref{eq:eta-of-metal}.
    \label{fig:BBH metal yield}
    }
\end{figure}

{In Fig.~\ref{fig:BBH metal yield} we show predictions for the yield as a function of metallicity under different physics variations from~\citet{2022ApJ...940..184V} in gray. {These predictions assume a solar metallicity value of $Z_\odot=0.0142$~\citep{2009ARA&A..47..481A}, the default value in \textsc{COMPAS}~\citep{2022ApJS..258...34R}, the remainder of our calculations therefore adopt this $Z_\odot$ value.} The flat behaviour at low metallicities followed by a steep drop off towards higher metallicities is characteristic for the BBH yield and has been found before; see, e.g., {Fig. 6 from~\citet{2018A&A...619A..77K}}, Fig. 1 from \citet{2019MNRAS.490.3740N}, {Fig. 1 from~\citet{2021MNRAS.502.4877S}}, Fig. 1 from \cite{2022MNRAS.516.5737B}, and Figs. 17 and 18 from \citet{2023MNRAS.524..426I}.}
Motivated by these predictions we take $dN_\mathrm{BBH}/dM_\mathrm{SF}(Z)$ to be of the form:
\begin{align}
\label{eq:eta-of-metal}
dN_\mathrm{BBH}&/dM_\mathrm{SF}(Z) = \nonumber \\ &y \ln(w - \log_{10}Z )\Theta(\log_{10}Z < w - 1),
\end{align}
so that $y$ sets the BBH yield at low metallicities, and as $\log_{10}Z$ approaches $w - 1$ from below, the BBH yield rapidly falls to zero.
Typical predictions have $10^{-5} < y < 6 \times 10^{-5}$ and $0.4 < w < 1.3$. To bound these possibilities, we consider a ``high yield" case with $(y, w)=(6 \times 10^{-5}, 1.3)$, a ``medium yield" case with $(y, w)=(3.5\times10^{-5},0.85)$ and a ``low yield" case with $(y, w) = (10^{-5}, 0.4)$, as shown by the red lines in Fig. \ref{fig:BBH metal yield}. 

\begin{figure}
    \centering
    \includegraphics{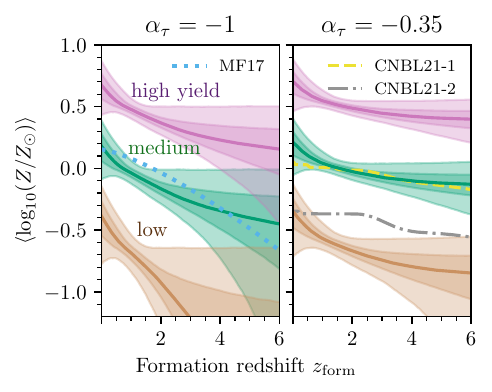}
    \caption{Inferred mean $\log_{10}$ metallicity as a function of redshift, assuming the default delay time distribution ($\alpha_\tau = -1$; left panel) or a shallow delay time distribution ($\alpha_\tau = -0.35$; right panel). We fix the standard deviation $\sigma_{\log_{10}}$ of the metallicity distribution to 0.2 dex at each redshift, and assume three different functions for the BBH yield as a function of metallicity: {high yield (pink), medium yield (green) and low yield (brown) as defined in Fig.~\ref{fig:BBH metal yield}. Solid lines show the median $\langle \log_{10}(Z / Z_\odot) \rangle$ at each $z_\mathrm{form}$ and shaded bands show 50\% and 90\% credible regions.} For comparison, we overplot three different measurements from the literature: \citealt{2017ApJ...840...39M} as the dotted, light blue curve (MF17), and two of the variations from \citealt{2021MNRAS.508.4994C} as the dashed, yellow curve (CNBL21-1) and dot-dashed gray curve (CNBL21-2).}
    \label{fig:meanZ}
\end{figure}

We approximate $p(Z \mid z_\mathrm{form})$ with a log-normal distribution with width $\sigma_{\log_{10}Z} = 0.2$ dex (though cf. \citealt{2022arXiv220610622C,2023ApJ...948..105V}; {we discuss the impact of the log-normal approximation in \S~\ref{sec:discussion}}). Using Eq.~\ref{eq:eta-of-redshift} with $dN_\mathrm{BBH}/dM_\mathrm{SF}(Z)$ given by Eq.~\ref{eq:eta-of-metal} and $dN_\mathrm{BBH}/dM_\mathrm{SF}(z_\mathrm{form})$ shown in Fig.~\ref{fig:yield-v-redshift}, we infer the mean log-metallicity $\langle\log_{10}(Z/Z_\odot)\rangle$ as a function of redshift. 

The results are shown in Fig.~\ref{fig:meanZ}. The left panel shows $\langle\log_{10}(Z/Z_\odot)\rangle$ inferred under the default $\alpha_\tau = -1$ delay time distribution, while the right panel shows the inference under a shallow $\alpha_\tau = -0.35$ delay time distribution. Each panel displays the inferred metallicity under the three different $dN_\mathrm{BBH}/dM_\mathrm{SF}(Z)$ assumptions: high yield (pink), medium yield (green) and low yield (brown). 

These measurements can be compared against various metallicity evolution results in the literature~\citep{2019A&ARv..27....3M,2022arXiv220610622C}. For reference, we overplot three examples: the mean log-metallicity from~\citet{2017ApJ...840...39M} (``MF17" or dotted, light blue line) and the peak metallicity curves from the left and right panels of Fig. 7 of~\citet{2021MNRAS.508.4994C}, calculated under two different assumptions (``CNBL21-1" and ``CNBL21-2" in the yellow dashed and gray dot-dashed lines, respectively). 
{While we can directly compare our results to~\citet{2017ApJ...840...39M} who also adopt a log-normal metallicity distribution, the comparison to~\citet{2021MNRAS.508.4994C} is less straightforward, since their metallicity distributions have an extended low-metallicity tail, resulting in a peak log-metallicity that is higher than the average. We expect this distinction to be less significant than the various uncertain assumptions illustrated in Fig.~\ref{fig:meanZ}, but we discuss the log-normal assumption further in \S\ref{sec:discussion}. We also note that our results assume a constant scatter in the log-metallicity of 0.2 dex. If we instead assumed a larger (smaller) scatter, our uncertainty in the inferred  $\langle \log_{10}(Z/Z_\odot) \rangle (z_\mathrm{form})$ would increase (decrease).}

Assuming the default delay time model with $\alpha_\tau = -1$ (left panel of Fig.~\ref{fig:meanZ}) and a medium BBH yield (green), we find a mean log-metallicity today of \result{$\langle \log_{10}(Z/Z_\odot) \rangle (z_\mathrm{form} = 0) = 0.2^{+0.2}_{-0.3}$}, and at redshift 4, \result{$\langle \log_{10}(Z/Z_\odot) \rangle (z_\mathrm{form} = 4) = -0.3^{+0.3}_{-0.4}$}. {(Here and in the following, we quote median values and 90\% credibility intervals.)} This is generally consistent with the~\citet{2017ApJ...840...39M} metallicity evolution (MF17) and the first peak metallicity curve from~\citet{2021MNRAS.508.4994C} (CNBL21-1), {although our average log-metallicity may evolve more steeply than these models over the well-constrained range $z_\mathrm{form} \lesssim 3$.\footnote{Note, however, that the steeper $\langle \log_{10}(Z/Z_\odot) \rangle (z_\mathrm{form})$ evolution may instead indicate a breakdown of the log-normal assumption; see \S\ref{sec:discussion}.}} If we assume a higher or lower BBH yield from Fig.~\ref{fig:BBH metal yield}, the inferred metallicities increase or decrease by a factor of $\sim 3$ (0.5 dex).

Meanwhile, different assumptions about the delay time distribution also affect the inferred metallicity evolution. A delay time model that favors longer delays, such as $\alpha_\tau = -0.35$, implies lower BBH yields across the plotted range $0 < z_\mathrm{form} < 6$ in order to avoid over-predicting the local merger rate, which largely consists of systems that formed at high redshifts. This lower BBH yield translates to higher inferred metallicities (right panel of Fig.~\ref{fig:meanZ}). For example, under the medium yield assumption and the $\alpha_\tau = -0.35$ delay time model, we infer \result{$\langle \log_{10}(Z/Z_\odot) \rangle (z_\mathrm{form} = 3) = -0.1^{+0.1}_{-0.1}$}. However, as discussed in the previous section, the $\alpha_\tau = -0.35$ provides a poor fit to the GW data compared to the $\alpha_\tau = -1$ model.

 Our results illustrates the promise of BBH mergers as probes of the cosmic metallicity evolution, {especially as GWs start providing improved constraints at formation redshifts past $z_\mathrm{form} > 3$ where the metallicity distribution inferred from other probes is highly uncertain~\citep[see, e.g., Fig. 7 in][]{2021MNRAS.508.4994C}}. Our inferred metallicity evolution is based on models of the delay time distribution, the BBH yield, and SFR.
 If we instead assumed that the metallicity evolution is known, we could use our measurement of $dN_\mathrm{BBH}/dM_\mathrm{SF}(z_\mathrm{form})$ from Fig.~\ref{fig:yield-v-redshift} and apply Eq.~\ref{eq:eta-of-redshift} to infer {the parameters governing the BBH yield as a function of metallicity}, similar to the analysis in~\citet{2021ApJ...914L..30F}. {In this study, the delay time distribution and BBH yield are inferred from population synthesis predictions, while we have used constraints from UV observations for the total SFR.  Although each of these components is plagued by uncertainties, our inference of the metallicity evolution can be seen as a consistency check between these different model components. We find that our default delay time distribution (from the ``All BBHs" prediction in Fig.~\ref{fig:delayTimes}), the medium yield assumption (dashed red line in Fig.~\ref{fig:BBH metal yield}) and the \citet{2021ApJ...919...88K} SFR imply a metallicity evolution that is consistent with external measurements in the literature~\citep[e.g.][]{2017ApJ...840...39M}. On the other hand, the high yield and low yield predictions, coupled with our other assumptions, require metallicities that are significantly different from the external measurements shown in Fig.~\ref{fig:meanZ} in order to match GW merger rates.  

\section{Oldest Black Holes in GWTC-3}
\label{sec:zform-pe}

\begin{figure}
    \includegraphics[]{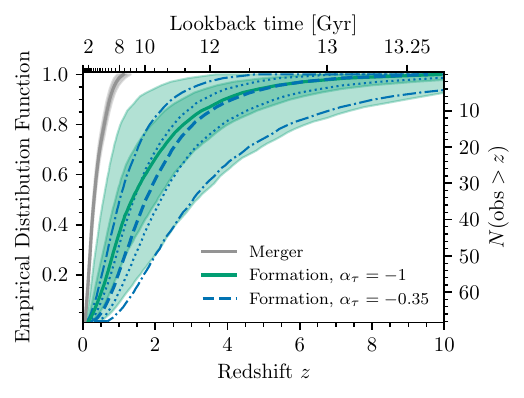}  \caption{\label{fig:EDF_zmerge_zform_GWTC3} Empirical distribution functions of the merger redshifts (gray) and formation redshifts (green and blue) of the 69 BBH events in GWTC-3 for fixed delay time distributions. The green filled band assumes a delay time distribution with power-law slope $\alpha_\tau$ = -1, and the blue, unfilled band assumes $\alpha_\tau = -0.35$. The solid green and dashed blue lines denote the median EDF, and bands enclose 50\% and 90\% credibility. Both delay time distributions assume $\tau_\mathrm{min} = 10$ Myr. The number of BBH events in GWTC-3 with a (merger or formation) redshift above a given value is denoted in the right-hand y-axis.}
\end{figure}

{Within a hierarchical Bayesian population analysis, we simultaneously infer the population properties and update our inference of individual-event parameters~\citep{2020ApJ...891L..31F,2020PhRvD.102h3026G,2020ApJ...895..128M,essick.fishbach.dcc,2021PhRvD.104h3008M}. Therefore, our fit to the BBH merger rate in the previous sections, modeled in terms of the progenitor formation rate and delay time distribution, allows us to jointly infer the probable merger redshift, delay time, and formation redshift for each BBH event in GWTC-3.}

{Figure~\ref{fig:EDF_zmerge_zform_GWTC3} shows the empirical distribution functions (EDFs) of the population-informed merger redshifts and formation redshifts of the 69 BBH events.} To construct the EDFs, we draw one sample from each of the 69 population-informed redshift ($z_\mathrm{merge}$ or $z_\mathrm{form}$) posteriors and order the samples from smallest to largest. 
The EDF at the $K$-th ordered $z$ sample takes the value $K/69$. Repeating this over 4000 draws, we calculate the median EDF of GWTC-3 redshifts and its 50\% and 90\% uncertainty bands. 

The gray band in Fig.~\ref{fig:EDF_zmerge_zform_GWTC3} shows the EDF of the merger redshifts $z_\mathrm{merge}$.
{We use the population-informed $z_\mathrm{merge}$ posteriors inferred under our default population model (with $\alpha_\tau = -1$), but the single-event $z_\mathrm{merge}$ posteriors are similar across the different models we consider, because they are largely informed by the GW data.}
The EDF of the formation redshifts $z_\mathrm{form}$ inferred under this default $\alpha_\tau = -1$ population model is plotted in green (filled bands) in Fig.~\ref{fig:EDF_zmerge_zform_GWTC3}, while the formation redshifts inferred under the $\alpha_\tau = -0.35$ model is shown in blue (unfilled bands). 
{The population-informed formation redshift inferred for each event depends on the assumed delay time distribution, which favors longer delay times for the $\alpha_\tau = -0.35$ model compared to the default $\alpha_\tau = -1$ model, together with the inferred formation rate, which acts as a prior over the formation redshifts (see Eq.~\ref{eq:ptf}).}
Under the $\alpha_\tau = -0.35$ delay time distribution, events tends to experience longer delay times and therefore have higher formation redshifts, {although the two EDFs overlap within statistical uncertainties due to the remaining uncertainty in the inferred formation histories}.

We can see from Fig.~\ref{fig:EDF_zmerge_zform_GWTC3} that although the observed BBH systems all merged at $z_\mathrm{merge} \lesssim 1$ (lookback times $t_L^\mathrm{merge} \lesssim 8$ Gyr), some of them experienced long delays of several Gyr between formation and merger according to either delay time distribution. 
The maximum merger redshift in GWTC-3 is \result{$\mathrm{max}(z_\mathrm{merge}) =1.1^{+0.2}_{-0.2}$} (median and 90\% credible interval).
Nevertheless, for our default delay time distribution, more than \result{$21$} events in GWTC-3 formed earlier than $z_\mathrm{form} > 1$, \result{$24^{+26}_{-20}$} events formed earlier than $z_\mathrm{form} > 2$ and \result{$13^{+24}_{-12}$} events formed earlier than $z_\mathrm{form} > 3$ (with 90\% credibility).
Furthermore, at least one event in GWTC-3 formed earlier than \result{$z_\mathrm{form} > 4.4$} (90\% credibility), when the Universe had only formed $\sim4\%$ of its stellar mass~\citep{2021ApJ...919...88K}.

\begin{figure}
    \includegraphics[]{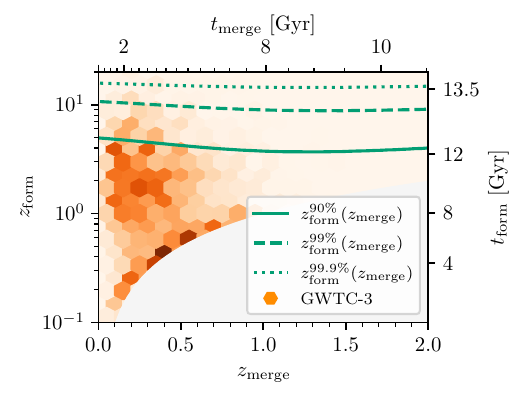}
    \caption{\label{fig:zform-zmerge-GWTC3} {\emph{Orange:} Joint distribution of the merger and formation redshifts of detected BBH events at GWTC-3 sensitivity, inferred under the default population model (with $\alpha_\tau = -1$ delay time distribution). We average over the population posterior uncertainty. Orange shading represents the probability density in a given $(z_\mathrm{merge}, z_\mathrm{form})$ hexagonal bin. For reference, top and right axes show corresponding lookback times. \emph{Green:} At each $z_\mathrm{merge}$, 90-, 99-, and 99.9-percentiles of the conditional $z_\mathrm{form}$ distribution (see Eq.~\ref{eq:ptf}). These represent the maximum formation redshift that we expect to probe given $\mathcal{O}(10)$, $\mathcal{O}(100)$ and $\mathcal{O}(1000)$ observations at a fixed $z_\mathrm{merge}$.}}
\end{figure}

{To understand which merger redshifts in GWTC-3 are responsible for the highest formation redshifts (i.e. the oldest BHs), we plot the joint distribution of merger and formation redshifts in Fig.~\ref{fig:zform-zmerge-GWTC3}. The orange histogram shows the detected distribution of $(z_\mathrm{merge}, z_\mathrm{form})$ according to our population inference under the default delay time model, with ``detected" referring to BBH events observable at GWTC-3 sensitivity. We average over the population uncertainty. This population-averaged, detected distribution is often referred to as a posterior population distribution (PPD) in the GW literature~\citep[e.g.][]{2019ApJ...882L..24A}.}

{Figure~\ref{fig:zform-zmerge-GWTC3} shows that, thanks to the delay time distribution that peaks at $\tau_\mathrm{min} = 10$ Myr, many BBH events experienced short delay times, leading to a region of high probability density at $z_\mathrm{form} \approx z_\mathrm{merge}$ in the detected distribution.
However, there is a second region of high density centered at $z_\mathrm{merge} \approx 0.2$, $z_\mathrm{form} \approx 2$.
This is caused by the fact that most observable BBH events merge at low redshifts $z_\mathrm{merge} \approx 0.2$, but we infer that the formation rate is very small at similarly low $z_\mathrm{form}$. Therefore, many of these low-redshift mergers experienced long delay times and formed at much higher redshifts.
This is broadly consistent with theoretical predictions for the formation redshifts of the observed, low-$z_\mathrm{merge}$ BBH systems~\citep{2016Natur.534..512B,2018MNRAS.481.5324M,2019MNRAS.487....2M,2021ApJ...907..110B}.
Our simplified ``backward population synthesis" approach therefore provides a straightforward test of whether predictions for the delay time distribution (which we fixed here to our default $\alpha_\tau = -1$, $\tau_\mathrm{min} = 10$ Myr model) yield reasonable BBH formation redshifts.
}

{Meanwhile, the green solid, dashed and dotted lines in Fig.~\ref{fig:zform-zmerge-GWTC3} show the 0.9, 0.99, and 0.999 quantiles of the conditional probability distribution $p(t_\mathrm{form} \mid t_\mathrm{merge})$ (given by Eq.~\ref{eq:ptf}), converting between lookback times to cosmological redshifts as necessary.} With $\mathcal{O}(N)$ events at a given merger redshift, we expect one of them to have formed at the $\frac{N - 1}{N}$ quantile of the $z_\mathrm{form}$ distribution. Therefore, we expect to probe the 0.9, 0.99, and 0.999 quantiles with $\mathcal{O}(10)$, $\mathcal{O}(100)$ and $\mathcal{O}(1000)$ events, respectively.
{Among the 69 BBH events in GWTC-3, the maximum formation redshift is $\mathrm{max}(z_\mathrm{form}) = 9.9^{+9.1}_{-6.2}$, which lies between the solid and dashed green lines.}
Interestingly, the largest $z_\mathrm{form}$ in a GW sample is unlikely to correspond to the largest $z_\mathrm{merge}$. 
As the GW detectors' sensitivity improves, we will probe higher formation redshifts $z_\mathrm{form}$ not directly because the detection horizon will increase to higher $z_\mathrm{merge}$, but because we will increase the BBH sample size, which allows us to probe higher quantiles of the $z_\mathrm{form}$ distribution. 
The highest $z_\mathrm{form}$ in the GW sample most likely corresponds to the $z_\mathrm{merge}$ that hosts the majority of detections ($z_\mathrm{merge}\approx0.2$ for GWTC-3). 

With thousands of observations at $ z_\mathrm{merge} < 1$ (to be expected within the next couple of observing runs) we expect to probe progenitor formation past $z_\mathrm{form} \gtrsim 15$ according to our default delay time model and our inferred progenitor formation history (see the dotted green line in Fig.~\ref{fig:zform-zmerge-GWTC3}). 
This suggests that with enough observing time, the current generation of GW observatories will already be sensitive to the first star formation in the Universe.
However, this prediction depends on the assumed delay time distribution.
This corroborates the need for next-generation GW detectors, which will provide a direct measurement of the merger rate at high redshifts.

\section{Discussion on simplifying assumptions}
\label{sec:discussion}
{
In this work, we have made several simplifying assumptions that can be relaxed in future work.
Here we briefly discuss the validity of these assumptions.

\paragraph{Mass and metallicity dependence of the delay time distribution}
In Figure \ref{fig:metals } we show the delay time distribution {from~\citet{2022ApJ...940..184V}}     
split between BBH systems formed at low (red), medium (yellow) and high (blue) metallicities. This shows a shift towards longer delay times for the highest metallicities.  
We suspect that this is caused by metallicity dependent winds, in particular during the Wolf-Rayet (WR) stage \citep[e.g.,][]{2005A&A...442..587V}. 
When the system consists of a WR + BH (the final stage before BBH formation), stronger winds at higher metallicities lead to wider separations, and thus longer delay times \citep[this was also concluded by e.g.,][]{2022A&A...665A..59B,2021MNRAS.505..663R}.

The metallicity dependence will cause the delay time distribution to correlate with the formation redshift $z_\mathrm{form}$, so that systems forming at high $z_\mathrm{form}$ likely experience shorter delays relative to systems forming at low $z_\mathrm{form}$. 
\maya{For a fixed BBH formation rate, this leads to a larger BBH merger rate at high redshifts compared to low redshifts, and therefore mimics the effect of changing the BBH formation rate. Neglecting this effect, we are likely overestimating the degree to which the BBH yield evolves with redshift. In other words, if we took into account the metallicity-dependence of the delay time distribution, we would infer a more gradual evolution of the BBH formation yield with redshift (Fig.~\ref{fig:yield-v-redshift}), and correspondingly a weaker dependence of metallicity on redshift (Fig.~\ref{fig:meanZ}). However, we expect the impact on our conclusions to be mild, because (a) the delay time distribution depends only weakly on metallicity over the relevant (low-metallicity) range for BBH formation, as seen by the small difference between the $Z\leq Z_\odot/10$ and $Z_\odot/10 < Z \leq Z_\odot/2$ delay time distributions in Fig.~\ref{fig:metals }; and (b) even neglecting the metallicity dependence of the delay time distribution, we infer relatively mild evolution of the average metallicity with redshift over the well-constrained range $0 < z_\mathrm{form} < 3$ in Fig.~\ref{fig:meanZ}.}

\begin{figure}
    \centering
    \includegraphics[width= 0.49\textwidth]{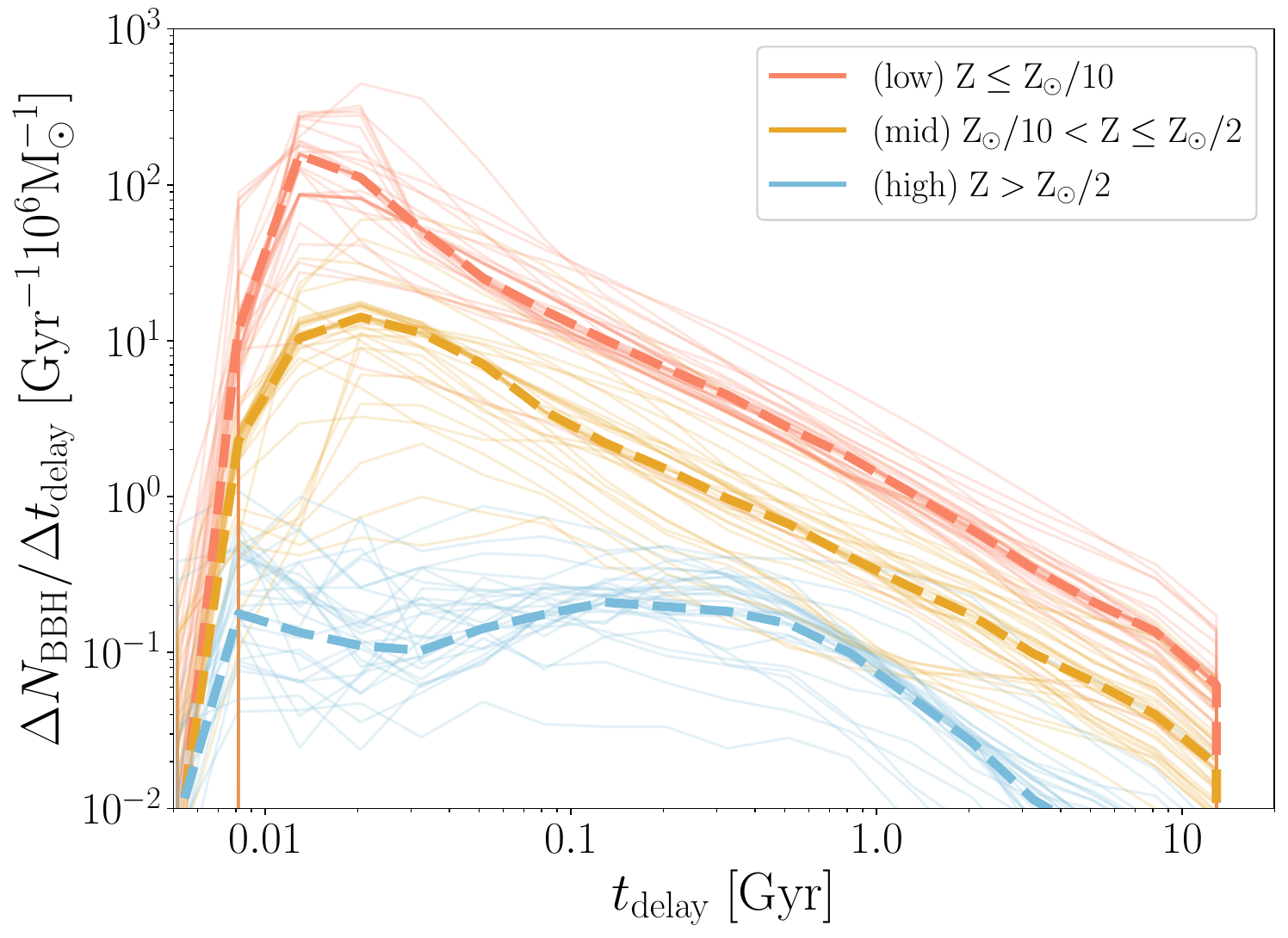} 
    \caption{\label{fig:metals } Similar to Figure \ref{fig:delayTimes}, but binned by formation metallicity. This shows that the delay time distribution shifts towards longer delay times for high metallicities. }
\end{figure}

{
The delay time distribution could furthermore depend on BBH mass. 
In particular, \cite{2022ApJ...931...17V} found that more massive BHs are formed exclusively by the stable mass transfer channel, while the common envelope channel dominates the formation of lower-mass systems. This result was confirmed by \citet{2022ApJ...935..126B} and \citet{2023MNRAS.520.5724B}. Given the distinct delay time distributions of these two channels (Fig. \ref{fig:delayTimes}), this implies a mass dependence of the delay time distribution. 
Currently, the contribution of each of these channels is an active area of research \citep{2019MNRAS.490.3740N, 2021A&A...647A.153B,2021A&A...650A.107M,2021ApJ...922..110G}. \maya{If lighter BBH systems tend to experience shorter delay times compared to the more massive systems, we expect a correlation between BBH mass and merger redshift, in which more massive BBH systems are more likely to merge at lower redshifts compared to low-mass BBH systems. With current GW data, there is no evidence for such a correlation, although it is not ruled out~\citep{2021ApJ...912...98F,2022ApJ...931...17V,2023PhRvX..13a1048A}. Because the evolution of the mass distribution with redshift is degenerate with the evolution of the overall merger rate, if a negative correlation between BBH mass and merger redshift exists, we are likely underestimating the BBH merger rate at high redshifts, which is dominated by light systems that are harder to detect. We may therefore also be underestimating the progenitor formation rate at high redshifts, although we expect this effect to be subdominant to other uncertainties given that GWTC-3 does not clearly display a correlation between BBH mass and merger redshift.}

In reality, the observed population of merging BBHs is most likely a mixture of multiple formation channels with a corresponding mixture of delay time distributions
\citep[e.g.,][]{2021ApJ...910..152Z,2021MNRAS.507.5224B,2022MNRAS.517.4034S,2023AAS...24147006G}.
Future work should simultaneously incorporate delay time distributions from several formation channels, including predictions from non-isolated binary evolution channels.
The mass (or spin; e.g.~\citealt{2022A&A...665A..59B}) dependence of the delay time distribution might also provide a solution to differentiate between formation channels: if a formation channel predicts unique observable properties (such as massive and highly spinning black holes), then our analysis can be repeated for a subset of the observations that meets these criteria. 
For example, \citet{2023MNRAS.522.5546F} leveraged predicted delay time distributions from dynamical assembly of BBHs in globular clusters to infer the globular cluster formation history from the observed merger rate of systems with misaligned spins, because misaligned spins indicate a possible dynamical origin.
%
}

\paragraph{Flexible progenitor formation rate models}
In this work, we used a simple three-parameter model for the progenitor formation rate, which follows a power-law in time at early times (high redshift) and exponentially decays at late times (low redshifts). Future work should allow for more flexible formation rate models, such as Gaussian processes~\citep{2019ApJ...886L...1V}, splines~\citep{2023ApJ...946...16E}, or autoregressive processes~\citep{2023arXiv230207289C}. In fact, there is already indication that our formation rate model provides a poor fit to the GWTC-3 data when we assume shallow delay time models ($\alpha_\tau = -0.7$ or $\alpha_\tau = -0.35$). This is consistent with the results of~\citet{2021ApJ...914L..30F}, who also found that short delay times were favored under the different models they considered for the BBH progenitor formation rate. However, because of the degeneracy between the delay time distribution and the progenitor formation rate, this may instead indicate that our formation rate parameterization fails to adequately fit the data. If the delay time distribution is indeed shallower than the default assumption of $\alpha_\tau = -1$, the formation rate likely deviates from our simple parameterization at formation redshifts $z_\mathrm{form} > 2$.

{
\paragraph{Deviations from a log-normal metallicity distribution}
When inferring the cosmic metallicity evolution in \S\ref{sec:metallicity}, we assumed a log-normal metallicity distribution at each redshift. However, there are observational and theoretical indications that the metallicity distribution may deviate from the log-normal approximation, featuring an extended low-metallicity tail that grows with redshift~\citep[see, e.g., the discussion in][]{2022arXiv220610622C}. Depending on the BBH yield as a function of metallicity (see Fig.~\ref{fig:BBH metal yield}), we expect GW observations to be predominantly sensitive to the amount of low-metallicity star formation as a function of redshift. 
Therefore, current GW data cannot distinguish between a log-normal metallicity distribution that peaks at lower metallicities and a distribution with an extended low-metallicity tail that peaks at higher metallicities. 
Future work should fit for more flexible metallicity distributions~\citep[e.g.][]{2023ApJ...948..105V} in order to account for this degeneracy in the location and shape of the metallicity distribution as a function of redshift.
}

\paragraph{Universality of the IMF}
Our calculations have assumed a universal \citet{2001MNRAS.322..231K} IMF. However, there are many indications that the IMF varies with metallicity \citep[see e.g., reviews by][]{2013pss5.book..115K,2018PASA...35...39H}. 
{The formation rate of BBH progenitors is unlikely to be strongly affected by variations in the high-redshift IMF within observational uncertainties, because the formation rate of high-mass stars, which is a combination of the SFR and the IMF, is better measured than the SFR or IMF independently~\citep{2018A&A...619A..77K,2020A&A...636A..10C}. Nevertheless, a non-universal IMF may affect the delay time distribution and the BBH mass distribution.}
Our current model only considers the delay time distribution, the BBH yield, and the SFR derived under a fixed IMF. Adding the IMF to this equation also implies that we can in principle use a similar method to measure the IMF at high redshift, in particular the transition between the IMF of Pop III stars and the IMF today. 


\section{Summary}
\label{sec:conclusion}

The preceding sections have demonstrated that existing GW observations by the LVK are starting to reveal the conditions of BBH formation beyond cosmic noon.
This argument stems from the fact that regardless of the precise formation scenario, some BBH systems are predicted to experience long delay times between the formation of their progenitor stars and their GW-driven merger. 
For this proof-of-principle study, we have adopted a few fixed delay time distributions motivated by predictions from isolated binary evolution. We then applied the assumed delay time distributions to GWTC-3 events to infer their progenitor formation redshifts, while simultaneously measuring the progenitor formation history out to $z_\mathrm{form} \gtrsim 4$ and implications for the cosmic metallicity evolution. Our main results are as follows:

\begin{enumerate}
    \item Fixing the delay time distribution to our default distribution with power-law slope $\alpha_\tau = -1$ and minimum delay of $\tau_\mathrm{min} = 10$ Myr, we infer the BBH progenitor formation rate as a function of redshift $z_\mathrm{form}$. {Assuming a star formation history model from~\citet{2021ApJ...919...88K}}, we find that the number of BBH progenitor systems formed per stellar mass was likely higher in the past than today (for our default delay time model, the BBH yield was \result{$6.4^{+9.4}_{-5.5}\times10^{-6}\,\Msun$ at $z_\mathrm{form} = 4$ compared to $0.134^{+1.6}_{-0.127}\times10^{-6}\,\Msun$ at $z_\mathrm{form} = 0$).}

    \item Combining our inferred BBH yield as a function of redshift with population synthesis predictions for the BBH yield as a function of metallicity, we measure the average metallicity as a function of redshift. {For our default delay time, BBH yield, and star formation history models}, we find that the mean log-metallicity $\langle \log_{10}(Z/Z_\odot) \rangle$ today is \result{$0.2^{+0.2}_{-0.3}$} and was \result{$-0.3^{+0.3}_{-0.4}$} at $z_\mathrm{form} = 4$. This is consistent with the metallicity evolution from~\citet{2017ApJ...840...39M}, highlighting that a simplified form of ``backward population synthesis"~\citep{2018ApJS..237....1A,2022arXiv220604062W} can provide a powerful self-consistency check on the various components of BBH population modeling.

    \item Our fit to the BBH population allows us to jointly infer the merger and formation redshifts of the 69 confident BBH events in GWTC-3. We find that GWTC-3 likely contains at least one BBH system that formed earlier than $z_\mathrm{form} > 4.4$.
\end{enumerate}

 The BBH merger rate is shaped by the SFR, IMF, BBH yield, and the delay time distribution. By fixing a few of these ingredients at a time and inferring the rest from the GW data, we can cross-check the various predictions of population synthesis with external measurements of the cosmic SFR and metallicity evolution. As the GW catalog grows and population synthesis simulations improve, the connection between these pieces will provide valuable insights into high-redshift and low-metallicity star formation, complementary to electromagnetic observations.

\section*{Software and Data}
{Posterior samples for the hyperparameters used in this work and data behind the figures are publicly available {on Zenodo~\citep{fishbach_maya_2023_8256745}}.
The population synthesis data used in this work is also available on Zenodo~\citep{l_van_son_2022_7080725,l_van_son_2022_7080164}.
This research made use of
Astropy \citep[\url{http://www.astropy.org}][]{2013A&A...558A..33A,Price-Whelan+2018,Astropy-Collaboration+2018, 2022ApJ...935..167A},
jupyter \citep[\url{https://jupyter.org}][]{jupyter},
matplotlib \citep[\url{https://matplotlib.org}][]{matplotlib},
numpy \citep[\url{https://numpy.org}][]{numpy}, 
scipy \citep[\url{https://scipy.org}][]{scipy}, jax \citep[\url{https://jax.readthedocs.io}][]{jax2018github}, and numpyro \citep[\url{https://num.pyro.ai}][]{phan2019composable,bingham2019pyro}.

\acknowledgments
We are grateful to an anonymous referee for pointing out a flaw in the calculations that appeared in an earlier draft of this work.
We furthermore thank Will Farr for his insightful comments on the manuscript and Reed Essick and Daniel Holz for inspiring discussions.
MF is grateful for the hospitality of Perimeter Institute where part of this work was carried out. Research at Perimeter Institute is supported in part by the Government of Canada through the Department of Innovation, Science and Economic Development Canada and by the Province of Ontario through the Ministry of Economic Development, Job Creation and Trade. 
LvS acknowledges partial financial support from the National Science Foundation under Grant No. (NSF grant number
2009131), the Netherlands Organisation for Scientific Research (NWO) as part of the Vidi research program BinWaves with project number 639.042.728 and the European Union’s Horizon 2020 research and innovation program from the European Research Council (ERC, Grant agreement No. 715063). This material is based upon work supported by NSF's LIGO Laboratory which is a major facility fully funded by the National Science Foundation. This is LIGO Document LIGO-P2300241.

\bibliographystyle{aasjournal}
\bibliography{references}

\appendix
\section{Details of Population Fit}
\label{sec:appendix}

In \S\ref{sec:Rf-inference}, we fit the population distribution of masses, spins and merger redshifts of the GWTC-3 BBH events. We use the 69 confident BBH events with false-alarm-rate FAR $<1$ yr$^{-1}$ and both component masses heavier than 3$\,M_\odot$, as selected by~\citet{2023PhRvX..13a1048A}. Our population model consists of independent distributions for component masses $m_1$ and $m_2$, effective inspiral spin parameter $\chi_\mathrm{eff}$, and redshift $z_\mathrm{merge}$ with corresponding hyperparameters $\Lambda_m$, $\Lambda_\chi$ and $\Lambda_z$. For the mass distribution $p(m_1, m_2 \mid \Lambda_m)$, we adopt the slightly modified version of the \textsc{Power Law + Peak} model~\citep{2018ApJ...856..173T} from~\citet{2023MNRAS.522.5546F}; see their Eqs.(A2)-(A6). For the $\chi_\mathrm{eff}$ distribution $p(\chi_\mathrm{eff} \mid \Lambda_\chi)$, we assume the \textsc{Truncated Gaussian} model, truncated to the physical range $-1 < \chi_\mathrm{eff} < 1$~\citep{2019MNRAS.484.4216R,2020ApJ...895..128M}. The merger redshift distribution is given by the formation rate $R_f(z_\mathrm{form})$ of Eq.~\ref{eq:Rf-KYZ} convolved with an assumed delay time distribution $p_d(\tau)$ of Eq.~\ref{eq:delay-pl}. Given hyperparameters $\Lambda_z$, the merger rate as a function of redshift is (see Eq.~\ref{eq:Rm-time}):
\begin{equation}
\label{eq:Rm}
R_m(z_\mathrm{merge} \mid \Lambda_z) = \int_{\tau_\mathrm{min}}^{\tau_\mathrm{max}} R_f(z_\mathrm{form}(z_\mathrm{merge}, \tau) \mid \Lambda_z) p_d(\tau) d\tau,
\end{equation}
where the formation redshift corresponding to a given merger redshift and delay time, $z_\mathrm{form}(z_\mathrm{merge}, \tau)$, is given by converting the lookback time $t_f = t_\mathrm{merge} + \tau$.
Fixing the delay time distribution $p(\tau)$, Eq.~\ref{eq:Rm} depends on the hyperparameters $\Lambda_z = \{\mathcal{N}$, $a$, $b$ \} from Eq.~\ref{eq:Rf-KYZ}. 
The merger rate is defined as the differential number density of mergers per comoving volume and source-frame time, $R_m \equiv dN/dV_cdt_s$. Converting this to a probability density over merger redshifts, i.e. the differential BBH number density per merger redshift and detector-frame time, we have:
\begin{equation}
    \frac{dN}{dz_\mathrm{merge}}\left(\Lambda_z\right) = N \frac{R_m(z_\mathrm{merge} \mid \Lambda_z) \frac{dV_c}{dz_\mathrm{merge}} (1 + z_\mathrm{merge})^{-1}}{\int_0^{z^\mathrm{max}_\mathrm{merge}} R_m(z_\mathrm{merge} \mid \Lambda_z) \frac{dV_c}{dz_\mathrm{merge}} (1 + z_\mathrm{merge})^{-1} dz_\mathrm{merge}}.
\end{equation}
The normalization $N$ in the above equation represents the total number of BBH sources that merge between redshift 0 and $z^\mathrm{max}_\mathrm{merge}$ in a given observation time. We are free to choose this maximum merger redshift as long as it is higher than the BBH horizon for GWTC-3; we choose $z^\mathrm{max}_\mathrm{merge} = 3$. Specifying this normalization $N$ is equivalent to specifying the amplitude $\mathcal{N}$ of the formation rate $R_f(z_\mathrm{form})$ of Eq.~\ref{eq:Rf-KYZ}. For consistency with other GW population analyses, we sample over $N$ in our population likelihood with a flat-in-log prior. 

The full population distribution in terms of source parameters $\theta = \{m_1, m_2, \chi_\mathrm{eff}, z_\mathrm{merge}\}$ and hyperparamters $\Lambda = \{\Lambda_m, \Lambda_\chi, \Lambda_z, N\}$ is:
\begin{equation}
\label{eq:full-pop-model}
    \frac{dN}{d\theta}(\Lambda) = \frac{dN}{dz_\mathrm{merge}}(\Lambda_z) p(m_1, m_2 \mid \Lambda_m)p(\chi_\mathrm{eff} \mid \Lambda_\chi),
\end{equation}
where $p(\ldots)$ denotes a normalized probability distribution.

We use an inhomogeneous Poisson process to model the likelihood for the GWTC-3 data consisting of $N_\mathrm{obs}$ independent observations $x = \{x_i\}_{i = 1}^{N_\mathrm{obs}}$ given hyperparameters $\Lambda$, marginalizing over source parameters $\theta$~\citep[see, e.g.][for reviews]{2004AIPC..735..195L,2019MNRAS.486.1086M,2019PASA...36...10T,2022hgwa.bookE..45V}:
\begin{equation}
\label{eq:pop-likelihood}
    p(x \mid \Lambda) = \exp\left(-\int \frac{dN}{d\theta}(\Lambda) P_\mathrm{det}(\theta) d\theta\right)\prod_{i = 1}^{N_\mathrm{obs}} \int p(x_i \mid \theta)\frac{dN}{d\theta}(\Lambda)d\theta
\end{equation}
We approximate both integrals in this likelihood with importance sampling (Monte Carlo) averages. 
For the first integral over $P_\mathrm{det}(\theta)d\theta$, we use the simulated signals from the GWTC-3 sensitivity estimates~\citep{ligo_scientific_collaboration_and_virgo_2021_5636816}, applying the appropriate weights to account for the simulated draw probability~\citep{2018CQGra..35n5009T,2022arXiv220400461E}. For the second integral over $p(x_i \mid \theta) d\theta$, we use the same GWTC-1, GWTC-2, GWTC-2.1 and GWTC-3 parameter estimation samples~\citep{2021SoftX..1300658A,10.5281_zenodo.5117703,ligo_scientific_collaboration_and_virgo_2021_5546663} used in the GWTC-3 population analysis~\citep{2023PhRvX..13a1048A}, applying the appropriate weights to account for the parameter estimation prior~\citep{2021arXiv210409508C}.

Recall that there is an additional integral in the likelihood of Eq.~\ref{eq:pop-likelihood}, because the population model $dN/d\theta$ is defined in terms of an integral over the delay time distribution $\tau$ (Eq.~\ref{eq:Rm}). However, $dN/d\theta$ always appears in the likelihood within another integral over $\theta$, either with $p(x \mid \theta)$ or $P_\mathrm{det}(\theta)$. For computational efficiency, we approximate both the integral over $\tau$ and the integral over $\theta$ with a single Monte Carlo integral over $m_1$, $m_2$, $\chi_\mathrm{eff}$ and $z_\mathrm{form}(z_\mathrm{merge}, \tau)$ samples. The $z_\mathrm{form}(z_\mathrm{merge}, \tau)$ samples are drawn by assigning a $\tau$ sample drawn from $p(\tau)$ to each of the $z_\mathrm{merge}$ samples. 

The posterior for the hyperparameters $\Lambda$ is given by the likelihood of Eq.~\ref{eq:pop-likelihood} with a choice of prior. We choose broad priors over all hyperparameters, given in Table~\ref{tab:GWprior}. We sample from the posterior with \textsc{numpyro}~\citep{phan2019composable,bingham2019pyro}. 

\begin{table}
    \centering
    \begin{tabular}{ l  l l l }
        \tableline
        {\bf BBH hyperparameter} & \textbf{Description} &  & \textbf{Prior} \\ \tableline\tableline
        $m_\mathrm{min} / M_\odot$ & Low-mass end of the primary mass spectrum &  & U(3, 12)\\
        $m_\mathrm{max} / M_\odot$ &  High-mass end of the primary mass spectrum &  & U(30, 80)\\
        $\eta_h$ &  Smoothing parameter for the low-mass end of the primary mass spectrum & & U(2, 20) \\
        $\eta_l$ &  Smoothing parameter for the high-mass end of the primary mass spectrum &  & U(2, 20)\\
        $\alpha$ & Power-law slope of the primary mass distribution &  & U(-6, -0.5) \\
        $f_\mathrm{peak}$ & Height of Gaussian peak in primary mass distribution &  & U(0.0001, 0.3) \\
        $m_\mathrm{peak}/M_\odot$ & Location of Gaussian peak &  & U(25, 50) \\
        $w_\mathrm{peak}/M_\odot$ & Width of Gaussian peak &  & U(2, 8) \\
        \tableline
        $\gamma$ & Power-law slope of the secondary mass distribution &  & U($-2$, $6$) \\
        \tableline
        $\mu$ & Center of $\chi_\mathrm{eff}$ distribution & & U(-0.5, 0.5) \\ 
        $\sigma$ & Width of $\chi_\mathrm{eff}$ distribution & & U(0.03, 0.5) \\
        \tableline
        $a$ & Power-law slope in $T(z_\mathrm{form})$ of the formation rate evolution & & U(0.3, 2.6) \\
        $b/\mathrm{Gyr}^{-1}$ & Characteristic timescale for exponential decline in $T(z_\mathrm{form})$ && U(0.0001, 0.001) \\
        \tableline
        $\log N$ & Normalization constant, total number of BBH mergers at $0 < z < 3$ & & U(7, 13) \\
        \tableline
    \end{tabular}
    \caption{
    Summary of hyperparameters $\Lambda$ describing the phenomenological BBH population model in Eq.~\ref{eq:full-pop-model}.  The notation U$(a, b)$ denotes a uniform distribution between $a$ and $b$.
    }
  \label{tab:GWprior}
\end{table}


\end{document}